\documentclass[acmsmall,screen]{acmart}

\usepackage{listings}
\usepackage{mathtools}

\DeclarePairedDelimiter\floor{\lfloor}{\rfloor}

\AtBeginDocument{%
  }

\setcopyright{none}
\copyrightyear{2025}
\acmYear{2025}
\acmDOI{}
\acmConference[Conference acronym 'XX]{Make sure to enter the correct
  conference title from your rights confirmation email}{June 03--05,
  2018}{Woodstock, NY}
\acmISBN{978-1-4503-XXXX-X/2018/06}

\usepackage{fancyhdr}

\begin{document}

\title{Encoding Numeric Computations and Infusing Heuristic Knowledge Using Integrity Constraints in stableKanren}

\author{Xiangyu Guo}
\email{Xiangyu.Guo@asu.edu}
\orcid{0000-0001-8120-2365}
\affiliation{%
  \institution{Arizona State University}
  \city{Tempe}
  \state{Arizona}
  \country{USA}
}

\author{Ajay Bansal}
\email{Ajay.Bansal@asu.edu}
\orcid{0000-0001-8639-5813}
\affiliation{%
  \institution{Arizona State University}
  \city{Tempe}
  \state{Arizona}
  \country{USA}
}

\renewcommand{\shortauthors}{Xiangyu Guo and Ajay Bansal}

\settopmatter{printacmref=false}
\settopmatter{printfolios=true}
\renewcommand\footnotetextcopyrightpermission[1]{}
\pagestyle{fancy}
\fancyfoot{}
\fancyfoot[R]{miniKanren'25}
\fancypagestyle{firstfancy}{
  \fancyhead{}
  \fancyhead[R]{miniKanren'25}
  \fancyfoot{}
}
\makeatletter
\let\@authorsaddresses\@empty
\makeatother

\begin{abstract}
	This paper presents examples of using integrity constraints in stableKanren to encode numeric computations for problem solving.
	Then, we use one of the examples to introduce multiple ways to infuse heuristic knowledge and reduce solving time.
	stableKanren is an extension of miniKanren that supports normal logic programs under stable model semantics.
	stableKanren further supports numeric computation by constructing a constraint store for integrity constraints.
	There are three ways to extend a relational programming language with numeric computations: relational number representation, grounding numbers to symbols, and constraint store construction.
	We demonstrate that the numeric computations in stableKanren have a straightforward numerical representation compared to relational number representations.
	More importantly, stableKanren balances symbolic and numeric computation in relational programming by avoiding the grounding of all numbers to symbols.
	Lastly, it also has simpler syntax compared to other constraint store construction approaches.
	stableKanren supports combinatorial search problem solving under a declarative generate and test paradigm.
	Such a paradigm generates all possible combinations of solutions to the problem, then applies a set of constraints to prune out the unwanted solutions.
	We demonstrate that different approaches to writing programs or queries affect the solver's performance in the SEND+MORE=MONEY puzzle.
	The performance gradually improves as more heuristic knowledge is infused through the programs or queries.
	Additionally, we show how to use an external function to achieve a hybrid solution.
\end{abstract}




\maketitle

\thispagestyle{firstfancy}

\section{Introduction}
\label{sec:introduction}

    We present two examples of writing numeric computations in a relational programming language, stableKanren.
    We then use the SEND+MORE=MONEY puzzle to introduce multiple ways of infusing heuristic knowledge to guide the internal search process in stableKanren.
    
    The foundation of relational programming is symbols, which are related through resolution and unification, resulting in a bidirectional relation rather than a unidirectional function.
    Friedman et al. built miniKanren to capture resolution and unification, the essence of Prolog, and show a natural way to extend functional programming to relational programming \cite{Friedman:2005:ReasonedSchemer1st}.
    The core miniKanren implementation introduces only a few operators to users: \textit{==} for \emph{unification}, \textit{fresh} for \emph{existential quantification}, \textit{conde} for \emph{disjunction}, and a \textit{run} interface for query.
    The \emph{conjunction} is captured by the sequential evaluation process naturally, so there is no operator for conjunction.
    However, the CPU is unidirectional, computing numbers as a function that takes inputs and produces an output.
    Therefore, adding numeric computation to a relational programming language is not straightforward.
    
    There are three approaches to add numeric computation.
    The first approach is to create bit-wise relations or a relational CPU and represent numbers as reversed bits, which is not easy to read and cannot utilize the capabilities of the numerical CPU.
    The second approach is to convert all numbers into symbols through grounding, which sacrifices top-down resolution and leaves a significant memory footprint.
    The third one is to construct a constraint store, where the numeric expressions obtain values along with the resolution, and evaluate the expressions once they have sufficient values.
    The constraint store approach creates a boundary between symbols and numbers, thereby balancing the number representation and memory usage.

    Under the constraint store approach, there are two types of constraints that can be introduced to relational programming languages: the \textit{variable constraints} and the \textit{integrity constraints} (Definition \ref{def:constraint-rule}).
    The variable constraints allow numeric constraints on relational variables; these constraints are applied to the variable during resolution and are evaluated once all variables have been unified with symbols.
    The integrity constraints control the outcome of a relational goal; these constraints are imposed on the goal and the relation between goals, and they are evaluated once sufficient goals are successfully proved by resolution.
    A goal is successfully proved through resolution, which also means the goal's variables are unified with symbols.
    Both variable and integrity constraints are able to represent whether a variable is unified or not unified with a symbol; hence, the relational programming language that has these constraints can encode the solutions to the combinatorial search problems like nqueens, graph coloring, hamiltonian cycles, etc \cite{Guo:2024:mkworkshop-stableKanren-CaVE, Alvis:2011:ckanren}.

    To add integrity constraints, the relational programming language also needs to have proper semantics for negation in the normal program (Definition \ref{def:normal-program}).
    Guo et al. extend miniKanren under stable model semantics \cite{Gelfond:1988:stable} to support negations in stableKanren \cite{Guo:2023:PPDP-stableKanren}.
    stableKanren also tracks integrity constraints through Constraints as Verifiers and Emitters CaVE (Definition \ref{def:emitter_verifier}) \cite{Guo:2024:mkworkshop-stableKanren-CaVE}.
    stableKanren introduces three more operators: \textit{noto} for \emph{negation}, \textit{defineo} for goal function definition, and \textit{constrainto} for controlling the goal function outcome.
    stableKanren allows combinatorial search problems to be solved in the declarative generate-and-test paradigm.


    In this paper, we demonstrate how to utilize integrity constraints in stableKanren to encode numeric computations.
    Furthermore, we introduce multiple ways of infusing heuristic knowledge into stableKanren, so that the internal search—the control part of the declarative algorithm—can be guided.
    We review the logic behind stableKanren and introduce the preliminaries in Section \ref{sec:preliminaries}.
    In Section \ref{sec:relatedwork}, we compare and contrast different approaches of adding numeric computations to relational programming languages.
    Then, we show two examples in Section \ref{sec:numericomputationsinsk} using the verifiers (Definition \ref{def:emitter_verifier}) in integrity constraints to perform numeric computations.
    We use query variables reordering (Section \ref{sec:qvr}), adding smaller constraints (Section \ref{sec:amc}), and critical emitters (Section \ref{sec:cm}) to reduce the running time to less than 20 seconds on the SEND+MORE=MONEY puzzle.
    Lastly, we present an external oracle verifier that leverages human heuristic knowledge to control the outcome of the \textit{assign} operation, thereby achieving hybrid solving.

\section{Preliminaries}
\label{sec:preliminaries}
    This section reviews several key logic programming definitions underlying stableKanren, including definite programs (Definition \ref{def:definite-program}), normal programs (Definition \ref{def:normal-program}), and integrity constraints (Definition \ref{def:constraint-rule}).
    The even negation cycles in the normal programs generate combinations, and the odd negation cycles prune out combinations \cite{Lin:2004:odd-and-even-cycles}.
    Integrity constraints are a special form of normal program clauses, in which they contain odd negation cycles.
    We present a logic program that generates multiple solutions and utilizes integrity constraints to eliminate unwanted solutions.
    Lastly, we use Constraints as Verifiers and Emitters (CaVE, Definition \ref{def:emitter_verifier}) to encode integrity constraints in stableKanren.

\subsection{Normal Programs, Integrity Constraints, and Generating Combinations}
\label{sec:integrityconstraints}
    Lloyd defines \textit{definite program}, \textit{normal program}, and \textit{integrity constraints} as follows \cite{Lloyd:1987:FoundationsOFLP}.

    \begin{definition}[definite program clause]
    \label{def:definite-program-clause}
    A \textit{definite program clause} is a clause of the form,
    \[A \leftarrow B_1, \cdots, B_n\]
    where $A, B_1, \dots , B_n$ are atoms\footnote{Atom is evaluated to be true or false.}.
    \end{definition}
    A definite program clause contains precisely one atom A in its consequent.
    $A$ is called the \textit{head} and $B_1, \dots , B_n$ is called the \textit{body} of the program clause.
    \begin{definition}[definite program]
    \label{def:definite-program}
    A \textit{definite program} is a finite set of definite program clauses.
    \end{definition}

    Based on the definition of the definite program clause, we have the definition of \textit{normal program clause} and \textit{normal program}.
    \begin{definition}[normal program clause]
    \label{def:normal-program-clause}
    A \textit{normal program clause} is a clause of the form,
    \[A \leftarrow B_1, \cdots , B_n, not \; B_{n+1}, \cdots , not \; B_{m}\]
    \end{definition}
    For a normal program clause, the body of a program clause is a conjunction of literals instead of atoms; $B_1, \cdots, B_n$ are \textit{positive literals} and $not \; B_{n+1}, \cdots, not \; B_{m}$ are \textit{negative literals}\footnote{A positive literal is just an atom, a negative literal is the negation of an atom.}.
    \begin{definition}[normal program]
    \label{def:normal-program}
    A \textit{normal program} is a finite set of normal program clauses.
    \end{definition}

    The \textit{integrity constraint} is defined as a headless normal program clause.
    \begin{definition}[integrity constraint]
    \label{def:constraint-rule}
    An \textit{integrity constraint} is a clause of the form,
    \[ \bot \leftarrow B_1, \cdots , B_n, not \; B_{n+1}, \cdots , not \; B_{m}\]
    where $B_1, \dots , B_m$ are atoms.
    \end{definition}
    For an integrity constraint, the head of a clause is $\bot$ (false) instead of an atom.
    So, the integrity constraints are headless.
    The constraint is violated only if all body literals are evaluated as $\top$ (true).
    When the clause's body is evaluated as $\top$, we get a formula that says true implies false.
    \[\bot \leftarrow \top\]
    This formula is impossible to prove true.
    So, the constraint is violated.
    The constraint can be satisfied if any literal in the body is evaluated to be false.
    Therefore, the integrity constraints mean that not all literals in the body can be proven true.

\subsection{Loops in Normal Programs, Generating and Eliminating Combinations}
\label{sec:loops}

    According to Van Gelder et al., there are sets of atoms named \emph{unfounded sets} in a normal program that can help us categorize the normal programs \cite{VanGelder:1991:well-founded-semantics}.
    \begin{definition}[unfounded set]
    \label{def:unfounded-set}
    Given a normal program, the atoms inside the unfounded set are only cyclically supporting each other, forming a loop.
    \end{definition}

    Replacing the atoms with literals in the unfounded set (loop) definition, we have informal definitions of \emph{positive loop} and \emph{negative loop}.
    \begin{definition}[positive loop]
    \label{def:positive-loop}
    A positive loop is a loop that contains no negative literals.
    \end{definition}
    \begin{definition}[negative loop]
    \label{def:negative-loop}
    A negative loop is a loop that has at least one negative literal involved.
    \end{definition}

    Using the number of negative literals, we can further define \emph{odd negative loop} and \emph{even negative loop} from Definition \ref{def:negative-loop}.
    \begin{definition}[odd negative loop]
    \label{def:odd-negative-loop}
    An odd negative loop is a loop that involves an odd number of negative literals.
    \end{definition}
    \begin{definition}[even negative loop]
    \label{def:even-negative-loop}
    An even negative loop is a loop that involves an even number of negative literals.
    \end{definition}
    
    An odd negative loop introduces a contradiction to the normal program.
    For example, the headless integrity constraint can be translated into an odd negative loop clause that creates a contradiction, as shown in Definition \ref{def:alter-constraint-rule}.
    \begin{definition}[alternative integrity constraint]
    \label{def:alter-constraint-rule}
    An \textit{alternative integrity constraint} is a clause of the form,
    \[ \boldsymbol{fail} \leftarrow B_1, \cdots , B_n, not \; B_{n+1}, \cdots , not \; B_{m}, \boldsymbol{not \; fail}\]
    \end{definition}
    This translation is equivalent to the original integrity constraint in Definition \ref{def:constraint-rule}, and no other transformations are needed.
    The \textit{fail} and \textit{not fail} pair introduces a contradiction when all other body literals $B_i$ are evaluated as true.

    Lin and Zhao point out that even negative loops in the normal program generate combinations, and the integrity constraints eliminate combinations \cite{Lin:2004:odd-and-even-cycles}.
    For example, given a propositional normal program (a program without variables) written in Prolog syntax \cite{ISO:1995:IIIe, ISO:2000:IIIf} as follows.
    \begin{lstlisting}
pick :- not free.
free :- not pick.
    \end{lstlisting}
    The program has one even negative loop.
    According to stable model semantics \cite{Gelfond:1988:stable}, either \textit{pick} or \textit{free} can be true, but they cannot be true or false at the same time.
    The truth value of \textit{pick} indicates one binary choice.
    The propositional example can be extended to a predicate version as follows.
    \begin{lstlisting}
pick(X) :- not free(X).
free(X) :- not pick(X).
    \end{lstlisting}
    The predicate program adds a variable, $X$, to increase expressiveness.
    However, the variable is \textit{free variable}; a program that contains free variables cannot be evaluated.
    Therefore, variables $X$ need to be grounded in values.
    The domain of the variable provided by a set of \textit{num} predicates is as follows.
    \begin{lstlisting}
num(1). num(2). num(3).
pick(X) :- num(X), not free(X).
free(X) :- num(X), not pick(X).
    \end{lstlisting}
    The above program grounds the variable $X$ to three values; therefore, it indicates three binary choices.
    In total, it generates $2^3 = 8$ possible combinations.
    
    A more complex example using two variables and an integrity constraint in Listing \ref{lst:normal},
    \begin{lstlisting}[caption=A logic program produces multiple models, label={lst:normal}]
num(1). num(2). num(3).
pick(X, Y) :- num(X), num(Y), not free(X, Y).
free(X, Y) :- num(X), num(Y), not pick(X, Y).
:- pick(X, Y), pick(U, V), X = U, Y != V.
    \end{lstlisting}
    The program describes a $3\times3$ board with a total of nine positions.
    The first line defines three numbers.
    Each position has either a pick or a non-pick choice.
    Therefore, it generates all possible combinations, resulting in $2^9 = 512$ models, under stable model semantics.
    The integrity constraint on the last line indicates that we do not pick the position in the same row.
    For example, \textit{pick(1, 1)} and \textit{pick(1, 2)} violate the integrity constraint.
    Therefore, the total number of models reduces from 512 to 64.

\subsection{Constraints as Verifiers and Emitters (CaVE) in stableKanren}
\label{sec:caveinsk}

    Constraints as Verifiers and Emitters
    CaVE (CaVE) categorizes the atoms $B_1 \cdots B_m$ in an integrity constraint (Definition \ref{def:constraint-rule}) into two types: an \textit{emitter} and a \textit{verifier}.
    \begin{definition}[emitter and verifier]
    \label{def:emitter_verifier}
        An integrity constraint can be written as \textit{emitters} and \textit{verifiers} in a clause of the form,
    \begin{align}
        \bot \leftarrow E_{1}, \cdots , E_{n}, not \; E_{n+1}, \cdots , not \; E_{m}, \\
    V_{1}, \cdots, V_{i}, not \; V_{i+1}, \cdots , not \; V_{j}
    \end{align}
    where $E_1, \dots , E_m$, $V_1$, \dots , $V_j$ are atoms.
    Specifically, $E_1, \dots , E_m$ are emitters.
    An \textit{emitter} is the head atom of a normal program clause that emits values.
    $V_1$, \dots , $V_j$ are verifiers.
    A \textit{verifier} is a boolean expression that receives and verifies values from the emitter.
    \end{definition}
    When resolution successfully unifies variables with values in the normal program clause, the values are emitted to the verifiers.
    All verifiers within the same integrity constraint are combined into a single constraint handler.
    \begin{definition}[constraint handler]
        A constraint handler is a concatenation (\textit{and} operator) of verifiers and $\top$ (true).
    \end{definition}
    Therefore, if all verifiers are true, the constraint handler returns true, indicating that a constraint has been violated.
    For example, the head atoms of the program in Listing \ref{lst:normal} are \textit{num}, \textit{pick}, and \textit{free}.
    The integrity constraint in Listing \ref{lst:normal} has two emitters, \textit{pick(X, Y)} and \textit{pick(U, V)}; two verifiers, \textit{X = U} and \textit{ Y != V} forming a constraint handler $(X = U) \land (Y != V) \land \top$.
    \begin{definition} [CaVE]
        CaVE breaks an integrity constraint into emitters and verifiers.
        The emitters are inserted as checkpoints in the resolution to emit values to verifiers.
        Once verifiers collect sufficient values, the constraint handler verifies the constraint; the verification result controls the resolution.
    \end{definition}
    During the resolution, the values emitted from \textit{pick} are captured by the constraint handler.
    For example, the \textit{pick} emits \textit{(1, 1)}, the partial constraint handler is $(1 = U) \land (1 != V) \land \top$.
    Later, \textit{pick} emits \textit{(1, 2)}, the previous parial constraint handler becomes $(1 = 1) \land (1 != 2) \land \top$, triggered a constraint violation.

    stableKanren uses CaVE to implement \textit{constrainto} \cite{Guo:2024:mkworkshop-stableKanren-CaVE}.
    So, the integrity constraints break into two parts in \textit{constrainto}.
    As a result, stableKanren supports solving combinatorial search problems using a declarative generate-and-test paradigm.
    For example, the logic program in Listing \ref{lst:normal} has the stableKanren equivalent as follows. 
    \begin{lstlisting}
(defineo (num x) (conde [(== 1 x)] [(== 2 x)] [(== 3 x)]))
(defineo (pick x y) (num x) (num y) (noto (free x y)))
(defineo (free x y) (num x) (num y) (noto (pick x y)))
(constrainto [(pick x y) (pick u v)] [(= x u) (not (= y v))])
    \end{lstlisting}
    The program has two parts: problem instances and a declarative algorithm.
    The first line is the problem instances \textit{nums}.
    The declarative algorithm also has two parts: generating and pruning.
    The \textit{pick} and \textit{free} form an even negation cycles that generate all combinations.
    And the \textit{constrainto} prunes out the unwanted combinations.

\section{Related Work}
\label{sec:relatedwork}
    This section reviews related work on adding numeric computation to relational programming languages.
    The challenge in adding numeric computation to a relational programming language is that symbolic unification only unifies symbols and lacks arithmetic operations.
    There are three ways to overcome the challenge.

\subsection{Relational Number Representation}
\label{sec:relationalnumberrepresentation}
    Friedman et al. propose a set of bit-wise relations to simulate a relational CPU \cite{Friedman:2005:ReasonedSchemer1st}.
    All numbers are represented as a list of reversed binary bits.
    For example, the number nineteen has the representation of (1 1 0 0 1), the first element in the list is the lowest binary bit.
    The unidirectional function \textit{xor} turns into a bidirectional relation \textit{xoro} as follows.
    \begin{lstlisting}
(define (xoro x y r)
  (conde [(== 0 x) (== 0 y) (== 0 r)] [(== 0 x) (== 1 y) (== 1 r)]
         [(== 1 x) (== 0 y) (== 1 r)] [(== 1 x) (== 1 y) (== 0 r)]))
    \end{lstlisting}
    The \textit{xoro} relation takes three parameters instead of two.
    This relation is maintained as long as all three parameters are successfully unified with their corresponding values.
    Similar relations are also defined for \textit{ando}, then using \textit{xoro} and \textit{ando} to define \textit{half-addero} and \textit{full-addero}.

\subsection{Converting to Symbols}
\label{sec:groundingtosymbols}
    Kaminski et al. design a bottom-up approach to solve logic programs \cite{Kaminski:2023:asp-grounding-foundations}.
    The bottom-up approach consists of two stages: grounding and problem-solving.
    All numeric operations are converted to symbols.
    Then the symbols are assigned a truth value to solve the Boolean equation. 
    For example, the integrity constraint we used to remove the same row in Listing \ref{lst:normal} produces 9 symbols from pick(1,1) to pick(1,3) and 18 propositional constraints after grounding.
    \begin{lstlisting}
:-pick(1,1),pick(1,2). :-pick(1,1),pick(1,3). :-pick(2,1),pick(2,2).
                        ...... 12 more ......
:-pick(2,3),pick(2,2). :-pick(3,3),pick(3,1). :-pick(3,3),pick(3,2).
    \end{lstlisting}
    Each symbol will be assigned a Boolean value to check whether the propositional constraints are violated or not.
    
\subsection{Constraint Store Construction}
\label{sec:constraintstoreconstruction}
    The last approach is to construct a constraint store to track all numeric expressions during resolution.
    Once the numeric expressions obtain sufficient values from unification, they can be evaluated.
    Alvis et al. implement cKanren where the numeric constraints are imposed on the variable by using the new operators \textit{infd}, \textit{plusfd}, and \textit{=/=fd} \cite{Alvis:2011:ckanren}.
    An example cKanren constraint is as follows.
    \begin{lstlisting}
(define (diago qi qj d rng)
  (fresh (qi+d qj+d)
    (infd qi+d qj+d rng)
    (plusfd qi d qi+d)
    (=/=fd qi+d qj)
    (plusfd qj d qj+d)
    (=/=fd qj+d qi)))
    \end{lstlisting}
    It ensures no two queens are placed on the diagonal for the nqueens problem.

    Guo et al. build an integrity constraint into stableKanren, where the numeric constraints are applied on the verifiers \cite{Guo:2024:mkworkshop-stableKanren-CaVE}.
    An example stableKanren constraint for the same nqueens problem is as follows.
    \begin{lstlisting}
(constrainto [(queen x y) (queen u v)]
             [(= (abs (- x u)) (abs (- y v)))
              (not (= x u)) (not (= y v))])
    \end{lstlisting}
    There are two emitters \textit{queen} emit two positions $(x,y)$, $(u,v)$ to the verifiers, and the verifiers perform the numeric computation to check the constraint.

\section{Numeric Computations in stableKanren}
\label{sec:numericomputationsinsk}

    The declarative generate-and-test paradigm reduces all problems to combinatorial search problems.
    All numbers and their possible relations are modeled as a combination search as well.
    The program generates all combinations of the relations between numbers, then prunes out invalid combinations.
    The verifiers (Definition \ref{def:emitter_verifier}) are the interface between symbols and numbers.
    As we mentioned at the end of Section \ref{sec:caveinsk}, the paradigm has two parts: problem instances and a declarative algorithm.
    In the following examples, the problem instances are always a sequence of numbers, depending on the actual problem size, and are omitted.
    We focus solely on the declarative algorithm.

\subsection{Arithmetic Relations}
\label{sec:arithmeticrelations}
    The arithmetic relations are bidirectional counterparts of the corresponding unidirectional functions.
    For example, the \textit{+} function in Scheme has a relation \textit{pluso} defined as follows.
    \begin{lstlisting}
(defineo (pluso x y z) (nums x) (nums y) (nums z) (noto (n_pluso x y z)))
(defineo (n_pluso x y z) (nums x) (nums y) (nums z) (noto (pluso x y z)))
(constrainto [(pluso x y z)] [(not (= (+ x y) z))])
(constrainto [(n_pluso x y z)] [(= (+ x y) z)])
    \end{lstlisting}
    The first two lines generate all possible combinations of the three numbers \((x, y, z)\).
    Some combinations do not meet the \textit{pluso} requirement, which requires $x + y = z$.
    The last two lines prune out these invalid combinations.

    The \textit{minuso} can be defined using \textit{pluso} by switching the parameter position.
    \begin{lstlisting}
(defineo (minuso z x y) (pluso x y z))
    \end{lstlisting}

    Similarly, \textit{multipo} and \textit{divido} are defined.
    The generating part is simple, so we only show the pruning part.
    \begin{lstlisting}
(constrainto [(multipo x y z)] [(not (= (* x y) z))])
(constrainto [(n_multipo x y z)] [(= (* x y) z)])

(constrainto [(divido x y q r)] [(or (= y 0) (>= r y)
                                     (not (= (+ (* y q) r) x)))])
(constrainto [(n_divido x y q r)] [(and (not (= y 0)) (< r y)
                                        (= (+ (* y q) r) x))])
    \end{lstlisting}
    The \textit{multipo} only requires $x*y=z$, but the \textit{divido} is slightly more complex, requiring the divisor not to be equal to 0, the remainder to be less than the divisor, in addition to $y*q + r =x$.

\subsection{SEND+MORE=MONEY}
\label{sec:sendmoremoney}
    The SEND + MORE = MONEY puzzle requires choosing different single-digit numbers (0-9) for each letter (SENDMORY) to satisfy the equation.
    An additional restriction is that there are no leading zeros in the numbers, so S and M can not be 0.
    The problem instances include \textit{letters} and \textit{values}, the \textit{letters} will have 8 symbols (SENDMORY), and the \textit{values} will have 10 symbols (0-9).
    The algorithm is shown in Listing \ref{lst:send-constraint}.
    \begin{lstlisting}[caption=SEND MORE MONEY puzzle in stableKanren with integrity constraints, label=lst:send-constraint, numbers=left]
(defineo (assign l v) (letters l) (values v) (noto (n_assign l v)))
(defineo (n_assign l v) (letters l) (values v) (noto (assign l v)))

(constrainto [(assign l1 v1) (assign l2 v2)]
             [(eq? l1 l2) (not (= v1 v2))])
(constrainto [(assign l1 v1) (assign l2 v2)]
             [(not (eq? l1 l2)) (= v1 v2)])

(defineo (assigned l) (fresh (v) (letters l) (values v) (assign l v)))
(constrainto [(letters l1) (noto (assigned l2))] [(eq? l1 l2)])

(constrainto [(assign 's s) (assign 'e e) (assign 'n n) (assign 'd d)
              (assign 'm m) (assign 'o o) (assign 'r r) (assign 'y y)]
             [(not (= (+ (* s 1000) (* e 100) (* n 10) (* d 1)
                         (* m 1000) (* o 100) (* r 10) (* e 1))
          (+ (* m 10000) (* o 1000) (* n 100) (* e 10) (* y 1))))])
    \end{lstlisting}
    The first two lines generate all possible combinations of the letters and values.
    The variables $l$ and $v$ in the \textit{assign} get values through \textit{letters} and \textit{values} as we have explained in Section \ref{sec:loops}.
    In lines 4 to 7, two constraints ensure that no letter takes more than one value and no value is assigned to more than one letter.
    Lines 9 and 10 ensure that each letter is assigned a unique value.
    Lines 12 to 16 use the puzzle's equation as a constraint to eliminate invalid value assignments.
    The symbols are emitted by the emitters \textit{assign} in lines 12 and 13; the verifiers receive and convert them to numbers so that the equation can be evaluated to verify the assignments in lines 14 to 16.

    A query uses the \textit{run} interface to produce an answer to the puzzle as follows.
    \begin{lstlisting}
> (run 1 (q) (fresh (s e n d m o r y)
               (assign 's s) (assign 'e e) (assign 'n n) (assign 'd d)
               (assign 'm m) (assign 'o o) (assign 'r r) (assign 'y y)
               (== q `(,s ,e ,n ,d ,m ,o ,r ,y))))
    \end{lstlisting}
    This query took 4,670 seconds to find one answer on a 2020 iMac with a 3.6 GHz Intel Core i9-10910 processor and 128GB of memory, which is not ideal for such a small puzzle.

\section{Infusing Heuristic Knowledge}
\label{sec:infuse}
    In this section, we use the SEND + MORE = MONEY puzzle as an example to illustrate different ways of incorporating heuristics into stableKanren.
    To satisfy the equation, the puzzle requires choosing different single-digit numbers (0-9) for each letter (SENDMORY).
    An additional restriction is that there are no leading zeros in the numbers, so S and M can not equal 0.
    We start from a basic, slow version of encoding in stableKanren in Section \ref{sec:sendmoremoney}.
    We gradually apply techniques such as query variable reordering, adding more constraints, and critical emitters.
    The solving time was drastically reduced to less than 20 seconds.
    Lastly, we demonstrate a hybrid solution that utilizes an external oracle verifier to further reduce the time.

\subsection{Variables Reordering}
\label{sec:qvr}
    Reordering the query variables is one way to control stableKanren's internal search.
    We measure the time spent on solving the puzzle with different starting letters.
    The ordering of the remaining seven letters is not specified in the query and is decided internally by the solver.
    Therefore, we have eight queries for each starting letter as follows.
    \begin{lstlisting}
> (run 1 (q) (assign 's q))
    \end{lstlisting}

    The running time in seconds is listed in Table \ref{tab:qvr}, along with the corresponding letter and value.
    \begin{table}[ht]
        \centering
        \begin{tabular}{|c|c|c|c|c|c|c|c|c|}\hline
             Letter&  s&  e&  n&  d&  m&  o&  r& y\\\hline
             Value&  9&  5&  6&  7&  1&  0&  8& 2\\\hline
             Time&  14242.16&  5730.59&  8721.58&  9778.90&  1052.88&  1080.75&  11074.77& 3717.25\\ \hline
        \end{tabular}
        \caption{Running time of different starting letters in the query}
        \label{tab:qvr}
    \end{table}

    As we can see, the running time ranges from 1080.75 (letter `o') to 14242.16 (letter `s') seconds.
    More importantly, the running time increases with the value ordering (0-9) in the \textit{values} goal function.
    The letter `o' is assigned with 0, so it prunes out the remaining search space (1-9) if the letter is placed at the first position in the query.
    In contrast, the letter `s' is assigned with 9, so it has to explore the entire search space if the letter is the first in the query.
    Therefore, if we reorder the query variables in the query based on the values they got assigned from 0 to 9.
    The query variables are ordered as OMYENDRS; the query only takes around 3 seconds to find one answer.
    However, if the query uses \textit{run*} to perform an exhaustive search, the running time exceeds 5000 seconds.
    The query variables sequence (OMYENDRS) was obtained after we knew the answer, but it shows the impact of the different ordering of query variables.

\subsection{Smaller Constraints}
\label{sec:amc}

    Replacing the big constraint with smaller constraints is another way to control stableKanren's internal search.
    The big constraint in Listing \ref{lst:send-constraint} requires all \textit{assign} relations to produce a value before verifying them.
    In the big constraint, the equation multiplies the value of each letter by the weight of each digit it represents.
    Therefore, it requires that all letters have a value.
    The equation can be represented as constraints on each digit.
    The equation breaks into addition on each digit and a carry value between digits.
    For example, the carry value on the least significant digit is 0, and the two digits are $D$ and $E$.
    We have $(D + E) \: mod \: 10 = Y$.
    And each digit creates a carry value to the next digit; the least significant bit creates $\floor{(D + E) \div 10}$ to the next digit.
    We break the big constraint into five smaller constraints as follows.
    \begin{lstlisting}
(constrainto [(assign 'd d) (assign 'e e) (assign 'y y)]
             [(not (= y (mod (+ d e) 10)))])

(constrainto [(assign 'd d) (assign 'e e) (assign 'n n) (assign 'r r)]
             [(not (= e (mod (+ n r (floor (/ (+ d e) 10))) 10)))])

(constrainto [(assign 'd d) (assign 'n n) (assign 'r r)
              (assign 'e e) (assign 'o o)]
             [(not (= n (mod (+ o e
                              (floor (/ (+ n r 
                              (floor (/ (+ d e) 10))) 10))) 10)))])

(constrainto [(assign 'n n) (assign 'd d) (assign 'r r) (assign 'e e)
              (assign 'o o) (assign 's s) (assign 'm m)]
             [(not (= o (mod (+ m s
                              (floor (/ (+ e o
                              (floor (/ (+ n r
                              (floor (/ (+ d e) 10))) 10))) 10))) 10)))])

(constrainto [(assign 'n n) (assign 'd d) (assign 'r r) (assign 'e e)
              (assign 'o o) (assign 's s) (assign 'm m)]
             [(not (= m (floor (/ (+ s m
                         (floor (/ (+ e o
                         (floor (/ (+ n r
                         (floor (/ (+ d e) 10))) 10))) 10))) 10))))])
    \end{lstlisting}
    Each constraint corresponds to a digit in the SEND+MORE=MONEY equation, from the lowest significant digit to the highest significant digit.
    The carry values pass through the digits as the computation moves from the lowest to the highest digit.
    So the number of emitters in each constraint gradually grows from 3, 4, 5 to 7.
    We measure the time performance on solving the puzzle with different starting letters using the same queries in Section \ref{sec:qvr}.
    The running time in seconds is listed in Table \ref{tab:mc}, along with the corresponding letter and value.
    \begin{table}[ht]
        \centering
        \begin{tabular}{|c|c|c|c|c|c|c|c|c|}\hline
             Letter&  s&  e&  n&  d&  m&  o&  r& y\\\hline
             Value&  9&  5&  6&  7&  1&  0&  8& 2\\\hline
             Time&  2708.52&  230.41&  332.39&  354.59&  44.26&  51.32&  118.26& 74.22\\ \hline
        \end{tabular}
        \caption{Running time of different starting letters in the query after adding more constraints}
        \label{tab:mc}
    \end{table}

    Compared to Table \ref{tab:qvr}, the performance improved 5 to 90 times after adding more constraints.
    The smaller constraints have fewer emitters; hence, the constraint can be verified earlier once sufficient values are collected.
    The query using the sequence OMYENDRS, introduced in Section \ref{sec:qvr}, takes only 0.46 seconds to find one answer, and 311.48 seconds if the query uses \textit{run*} to perform an exhaustive search.

\subsection{Critical Emitters}
\label{sec:cm}
    In Section \ref{sec:qvr}, the query variables sequence was created by knowing the answers, but it showed that the ordering of the query variables affects the internal search space.
    With additional constraints introduced in Section \ref{sec:amc}, we can refine the query variable sequence by identifying critical emitters.
    An emitter is critical if future emitters depend on the outcome of the current one.
    In the SEND+MORE=MONEY example, the least significant digits are more critical than the most significant digits, since the carry values are generated by the lower digits and passed through to the higher digits.
    So, the sequence from the lowest digits to the highest digits will be YDE, ENR, NEO, OSM, and M.
    Combining them and removing the duplicates, the query variables sequence is YDENROSM.
    An exhaustive query using \textit{run*} under the sequence YDENROSM only takes 15.69 seconds to produce the answer.

\subsection{External Oracle Verifier}
\label{sec:ov}
    Solving from the lowest digits to the highest digits is not the easiest path in the puzzle.
    It is easier to find a partial answer starting from the highest digits.
    From $S+M+C_0=MO$, the biggest possible outcome to the sum of two single digits $S$ and $M$ is 18; even with a potential carry $C_0$ from the lower digits, the sum does not exceed 19.
    Therefore, $M$ must be 1.
    Now the equation is $S+1+C_0=1O$, $C_0$ is either 0 or 1 from the lower digits, so $S\ge8$ to produce $1O$.
    Hence, $O$ can either be 0 or 1.
    Since $M=1$, and no letter is taking the same value, $O$ must be 0, the equation turns into $S+1+C_0=10$.
    If $C_0=1$ from the lower digits, $S=8$; otherwise $S=9$.
    In the middle digits $E+0+C_1=C_0N$ to produce $C_0=1$, $E$ needs to be 9, and the carry from lower digits $C_1=1$. 
    So, $N$ needs to be 0, which conflicts with $O=0$.
    The middle digits $E+0+C_1=C_0N$ cannot produce $C_0=1$, $C_0$ must be 0, so $S=9$.
    We can create an oracle verifier function that contains the above human reasoning, but we simplified it to storing the partial answer.
    \begin{lstlisting}
(define (oracle l v)
  (or (and (eq? l 's) (= v 9))
      (and (eq? l 'm) (= v 1))
      (and (eq? l 'o) (= v 0))
      (or (eq? l 'e) (eq? l 'n) (eq? l 'd) (eq? l 'r) (eq? l 'y))))

(constrainto [(assign l v)] [(not (oracle l v))])
    \end{lstlisting}
    Whenever the emitter \textit{assign} finds an assignment to a letter, the oracle verifier checks the value.
    This hybrid solving approach reduces the time required to query the YDENROSM sequence to 5.17 seconds.
    If the highest digits OSM are moved to the first, the sequence becomes OSMYDENRO.
    It only takes 1.89 seconds to complete an exhaustive search.

\section{Conclusion and Future Work}
\label{sec:conclusion}
    In conclusion, this paper shows the advantages of using integrity constraints (Definition \ref{def:constraint-rule}) in stableKanren to encode numeric computations.
    Unlike creating bitwise operators and representing relational numbers in bits, the numeric computations in stableKanren are easy to understand and utilize the numeric capabilities of the CPU.
    Unlike grounding all numbers to symbols and removing the top-down resolution, the numeric computations in stableKanren remain top-down and only track the necessary numbers, constraints, and expressions during resolution.
    Unlike constraining variables and controlling the algorithm imperatively, the numeric computations in stableKanren use integrity constraints to transform symbols to numbers in verifiers and declaratively control the outcome of the goals.
    
    All examples we showed are combinatorial search problems in stableKanren.
    The declarative algorithm only provides the logic of the solution, without providing any control.
    The solver controls the solving process.
    The verifiers (Definition \ref{def:emitter_verifier}) are the interface between symbolic and numeric computations.
    Currently, stableKanren has no heuristic; it only uses brute force.
    The solving speed depends on how the constraints are written.
    This paper also demonstrates how to integrate heuristic knowledge into stableKanren by reordering query variables, adding constraints, and utilizing critical emitters.
    Moreover, we illustrate using an external oracle verifier to achieve hybrid solving.

    There are several areas for future improvement.
    The verifier only checks the outcome but provides no guidance on the resolution.
    Adding a generic heuristic function could improve the performance for some problems.
    The SQL engine generates and optimizes a query plan before executing the actual query.
    The stableKanren solver can have a built-in optimizer to reorder the query variables, identify critical emitters, and analyze constraints, as we did in Section \ref{sec:infuse}.
    For example, the critical emitter analysis can be performed by topological sorting.
    The Conflict Driven Nogood Learning (CDNL) algorithm is the state-of-the-art algorithm for bottom-up solving \cite{Gebser:2007:conflict-driven-answer-solving}.
    The stablekanren solver can have a top-down version of CDNL to add generic heuristic capability.
    More specifically, when the symbols are numbers, constraint logic programming (CLP) techniques can be used for improving solving speed.
    The SEND+MORE=MONEY puzzle can be solved by cKanren in under a second.
    The stableKanren solver can have built-in CLP techniques and use them to prune the resolution search space.
    We believe that there is no unified heuristic that works for all NP-hard problems.
    Therefore, the stableKanren should not only provide internal, built-in heuristics but also allow external domain-specific heuristics for problem-solving.
    Moreover, the size of the problem instances varies; every change requires redefining instances in the program.
    So, a high-order relation that can generate instances based on the given number $n$ can reduce the burden.
    Lastly, the emitter only emits bound values, making it cannot create an application like a relational interpreter.
    Hence, allowing the emitter to emit unbound variables and the verifier to leave a placeholder to verify later will be helpful.


\bibliographystyle{ACM-Reference-Format}
\bibliography{ref}


\end{document}